hep-th/9403091

\def\H{{\cal H}}
\def\Ht{{\tilde H}}

\def\N{{\cal N}}
\def\dphi{{\delta\varphi}}

\def\nubble{{\dot{\N}\over \N}}
\def\dgij{{\delta\gamma_{ij}}}
\def\dpij{{\delta\pi^{ij}}}
\def\dN{{\delta\N}}
\def\dNi{{\delta\N^i}}
\def\dpfi{{\delta\pi^A_{\varphi}}}

\def\B{{\,^{\scriptscriptstyle (s)}B}}

\def\D{{\,^{\scriptscriptstyle (v)}D}}
\def\V{{\,^{\scriptscriptstyle (v)}V}}
\def\E{{\,^{\scriptscriptstyle (s)}E}}
\def\h{{\,^{\scriptscriptstyle (t)}h}}
\def\spsi{{\,^{\scriptscriptstyle (s)}\psi}}
\def\1s{{\,^{\scriptscriptstyle (s1)}\pi}}
\def\2s{{\,^{\scriptscriptstyle (s2)}\pi}}
\def\vpi{{\,^{\scriptscriptstyle (v)}\pi}}
\def\tpi{{\,^{\scriptscriptstyle (t)}\pi}}

\documentstyle [11pt]{article}    % Specifies the document style.

\title{Path Integral Quantization of Cosmological Perturbations}

\author{S. Anderegg and V. Mukhanov \\ Institut f\"{u}r Theoretische Physik \\
Eidgen\"{o}ssische Technische Hochschule \\ H\"{o}nggerberg, CH-8093
Z\"{u}rich, Switzerland}

\date{ETH-TH/94-09, March 1994}   % Deleting this command produces today's
\begin{document}           % End of preamble and beginning of text.

\maketitle

\begin{abstract}
We derive the first order canonical formulation of cosmological perturbation
theory  in a Universe filled by a few scalar fields. This theory is quantized
via well-defined Hamiltonian path integral. The propagator which describes the
evolution of the initial (for instance, vacuum) state, is calculated.
\end{abstract}

\section{Introduction}

The quantization of linearized matter and metric perturbations in the Friedmann
Universe has become an especially interesting topic in connection with
inflationary scenarios of the Universe evolution.

Actually, if the Universe went through an inflationary stage, then the initial
quantum fluctuations can explain the observable large-scale structure of it
(see, for instance, \cite{MFB92}).

One of the simplest models for inflation is the chaotic inflation when the
Universe is filled by homogeneous scalar fields \cite{Linde}. In this model,
small inhomogeneities in the distribution of scalar fields create  the metric
perturbations. In addition there can be fluctuations of the metric which are
not due to inhomogeneities of the matter (gravitational waves).

The purpose of this paper is to present a selfconsistent quantum theory of
these perturbations starting from first principles.

It was shown in \cite{JETP88} that the Lagrangean for the scalar cosmological
perturbations derived via expansion of the Einstein-Hilbert action can be
expressed entirely in terms of one gauge-invariant variable which describes the
collective degree of freedom of the metric and matter perturbations. This
reduced Lagrangean was then the starting point for the quantization.

There are several disadvantages in the above mentioned semi-quantum approach.
First, it does not permit us to reveal the connection between fundamental
quantum theory of gravity and quantization of small metric perturbations. In
particular, the role of gauge group (diffeomorphisms) is not clear there.
Second, to reduce the action to the function of only one variable, we need to
use some of the Einstein equations.

It is much more easy to clarify the role of these equations in the canonical
quantum theory, where they are just algebraic constraints. Further, the
calculations in Lagrangean theory are not very straightforward and it is a
tricky point to generalize them to include in the consideration several scalar
fields \cite{Deruelle, Langlois}.

And finally, the Lagrangean theory is not very suited to relate the functional
approach to quantization of perturbations with "fundamental" canonical or path
integral quantum gravity where the question about "boundary conditions" for a
quantized Universe is well posed \cite{Hal_Haw,Hartle_Haw}.

Our method in this paper is to expand the canonical ADM action for gravity and
$N$ scalar fields up to second order in the perturbation variables. Thus the
first order formulation of the classical theory of cosmological perturbations
is derived. Then we quantize this theory via path integral in the Hamiltonian
formalism where the measure is well defined. As a result the integration over
the (infinite) volume of group (diffeomorphisms) is factorized explicitly and
the problem is reduced to the quantization of a set of gauge invariant fields
with time-dependent masses. The propagators and evolution of initial vacuum
state are calculated. The details of calculations will be presented in a
forecoming paper \cite{And_Mukh}.

We use the units in which $c=\hbar = 16\pi G = 1$ and adopt MTW conventions
\cite{MTW}.

\section{Action}

If we write the metric in the ADM form \cite{ADM},
\begin{equation}
ds^2 = -(\N^2 - \N_i\N^i)dt^2 + 2\N_i dx^i dt + \gamma_{ij} dx^i dx^j,
\label{eq:backmetric}
\end{equation}
then the first order action for gravity and $N$-scalar fields $\varphi_A$
($A=1,\cdots ,N$) with potentials $V^A(\varphi)$ is \cite{Hal_Haw, DeWitt}:
\begin{equation}
S = \int d^3xdt(\pi^{ij}\dot{\gamma}_{ij} + \sum_A \pi^A_\varphi
\dot{\varphi}_A- \N^\alpha \H_\alpha), \label{eq:ADMaction}
\end{equation}
Here, the dot means the derivative with respect to coordinate time $t$,
$\pi^{ij}$ and $\pi^A_\varphi$ are the momenta conjugated to $\gamma_{ij} $ and
$\varphi_A$. The lapse $\N(x,t)$ and the shift $\N^i(x,t)$ functions play the
role of Lagrangean multipliers.

Correspondingly, the four constraints are
\begin{eqnarray}
\H_0 & = & G_{ijkl}\pi^{ij}\pi^{kl} - \sqrt{\gamma}~^{(3)}R +
\label{eq:scalconst} \\
 & & \mbox{} + \sum_A \biggl( {(\pi^A_{\varphi})^2\over {2\sqrt{\gamma}}} +
{\sqrt{\gamma} \over 2}\gamma^{ij} \varphi^A_{,i} \varphi^A_{,j} + \sqrt\gamma
V^A(\varphi) \biggr) \nonumber \\
\H_i & = & -2\gamma_{ij}\pi^{jk}_{~~|k} + \sum_A\pi_\varphi^A \varphi^A_{,i}
\label{eq:vecconst}
\end{eqnarray}
where
\begin{equation}
G_{ijkl}=
{1\over{2\sqrt{\gamma}}}(\gamma_{ik}\gamma_{jl}+\gamma_{il}\gamma_{jk}-\gamma_{ij}\gamma_{kl})~,\qquad \gamma = \mbox{det}(\gamma_{ij})
\end{equation}
and the bar $|$ denotes covariant derivatives with respect to the three-metric
$\gamma_{ij}$.

\section{Background}

In an isotropic, flat\footnote{The results can easily be generalized for closed
and open Universes.} Universe, the coordinate system can be choosen in such a
manner that
\begin{equation}
\begin{array}{rclrcl}
\varphi_A & = & \varphi_A(t), & \pi_\varphi & = & \pi_\varphi^A (t), \nonumber
\\
\gamma_{ij}  & = &  a^2(t)\delta_{ij}, &  \pi_\gamma^{ij} & = & \delta_{ij}
\pi_a/6a \label{eq:hom_solution} \\
\N & = & \N(t), & \N^i & = & 0 .\nonumber
\end{array}
\end{equation}

Then the action~(\ref{eq:ADMaction}) reduces to
\begin{equation}
S = \int dtd^3x \biggl(\pi_a \dot{a} + \sum_A \pi^A_\varphi \dot{\varphi}_A -
\N\H_0 \biggr), \label{eq:backation2}
\end{equation}
where now the superhamiltonian
\begin{equation}
\H_0 = {- (\pi_a)^2 \over{24a}} + \sum_A \biggl({(\pi ^A_\varphi)^2
\over{2a^3}} + a^3 V^A(\varphi) \biggr), \label{eq:Backhamiltonian}
\end{equation}
is constrained to vanish. Using Hamilton eqs. we can express the momenta
$\pi_a$ and $\pi^A_\varphi$ in terms of the "velocities" $\dot{a}$ and
$\dot{\varphi}^A$:
\begin{equation}
\pi_a = -12{a \dot{a}\over{\N}}, \qquad \pi_\varphi^A =
{a^3\over{\N}}\dot{\varphi}_A. \label{eq:velocities}
\end{equation}
{}From the vanishing Hamiltonian~(\ref{eq:Backhamiltonian}) and other Hamilton
eqs. with taking into account~(\ref{eq:velocities}), one obtains the the
following eqs. for the background Friedmann Universe:
\begin{equation}
\ddot{\varphi}_A + (3H - \nubble) \dot{\varphi}_A  + \N^2 V,^A_{\varphi} = 0,
\label{KGeqn}
\end{equation}
\begin{equation}
6H^2 = \sum_A \biggl({1\over{2}} (\dot{\varphi}_A)^2 + \N^2 V^A(\varphi)
\biggr),  \label{eq:background}
\end{equation}
where $H \equiv \dot{a}/a$ coincides with the Hubble parameter only if we
choose the gauge $\N(t) = 1$. Later, these eqs. will heavily be used to
simplify the action for cosmological perturbations.

\section{Perturbations}

To study the behaviour of perturbations in a flat Universe we consider small
deviations from the homogeneous solutions~(\ref{eq:hom_solution}), such that
for the $3$-scalars one has

\begin{eqnarray}
\varphi^A(x,t)  & = &  \varphi^A(t) + \dphi^A(x,t) \nonumber \\
\pi^A_\varphi(x,t) & = & \pi_\varphi^A(t) + \dpfi(x,t) \\
\N(x,t) & = & \N(t) + \dN(x,t)   \nonumber
\end{eqnarray}
It is convenient to write the perturbations in $3$-vector $\dNi$ and $3$-tensor
$\dgij$ and $\dpij$ as a sum of scalar $(s)$, vector $(v)$ and tensor $(t)$
peaces which don't interfere in the linear approximation:
\begin{eqnarray}
\dNi & = & \B^{,i} + \D^i  \nonumber \\
\dgij & = & a^2[- 2 \spsi \delta_{ij} + 2 \E_{,ij} + \V_{i,j} + \V_{j,i} +
\h_{ij}] \nonumber \\
\dpij & = & {1\over a^2}[ \1s \delta^{ij} + \2s^{,ij} + \vpi^{i,j} + \vpi^{j,i}
+ \tpi^{ij}]
\end{eqnarray}
where the comma means usual derivative with respect to corresponding spatial
coordinates. The vectors $D^i,V_i, \pi^i$ and tensors $\h_{ij},\tpi^{ij}$
satisfy the following conditions:\footnote{Here and in  what follows, we omit
repeating the superscripts $(s)$ and $(v)$ whenever it is clear which of the
decoupled peaces is considerd.}
\begin{equation}
\begin{array}{lrc}
D^i_{~,i} = 0 ,&  V_i^{~,i} = 0 , & \h ^i_{~i} = 0 = \h^i_{~k,i} \\
              & \vpi^i_{,i} = 0 , & \tpi^i{~_i} = 0 = \tpi^i_{~k,i}
\end{array}
\end{equation}
We will raise and lower indices by the unit tensor $\delta_{ij}$.

Expanding the action~(\ref{eq:ADMaction}) in perturbations, we find that the
first order terms vanish on the eqs. of motion for the background model whereas
the second order terms give the action for the perturbations.

This action consists of three decoupled terms which correspond  to {\it
scalar}, {\it vector} and {\it tensor} (gravitational waves) perturbations:
\begin{equation}
\begin{array}{lll}
\delta_2 S & =  &\,^{\scriptscriptstyle (s)}S(\dphi^A,\dpfi, \psi, E, \1s, \2s)
  \\
           &  & + \,^{\scriptscriptstyle (v)}S(V_i,\vpi^i) +
\,^{\scriptscriptstyle (t)}S(\h_{ij},\tpi^{ij})
\end{array}
\end{equation}
The explicit form of the action for {\bf scalar perturbations} up to total
derivatives is
\begin{equation}
\,^{\scriptscriptstyle (s)}S = \int dt d^3x \biggl(\pi_{\psi} \dot{\psi} +\pi_F
\dot{F} + \sum_A \dpfi \delta \dot{\varphi}_A - \,^{\scriptscriptstyle
(s)}\lambda^0 \,^{\scriptscriptstyle (s)}C_0 - \,^{\scriptscriptstyle
(s)}\lambda^1 \,^{\scriptscriptstyle (s)}C_1 - \,^{\scriptscriptstyle (s)}\H
\biggr) \label{scalaraction}
\end{equation}
where the two constraints linear in momenta are
\begin{eqnarray}
^{\scriptscriptstyle (s)}C_0 & = &  - {H\over{\N}}\pi_\psi -
{4a^3\over{\N^2}}(H^2+\dot{H}-H\nubble)(3\psi - F) - 4a\Delta \psi \nonumber \\
 &   &  \mbox{} + \sum_A \biggl({\dot{\varphi}^A \over \N} \dpfi + a^3
V^A_{,\varphi}\dphi_A\biggr), \label{scalarconstr} \\
^{\scriptscriptstyle (s)}C_1 & = &  \pi_F -{4a^3 H\over{\N}}(\psi + F)  -
\sum_A {a^3 \dot{\varphi}^A\over \N} \dphi_A, \nonumber
\end{eqnarray}
and the Hamiltonian for scalar perturbations $\,^{\scriptscriptstyle (s)}\H$
takes the form
\begin{eqnarray}
\,^{\scriptscriptstyle (s)}\H & = & {\N \over{8a^3}}(3 \pi_F^2 +2 \pi_F
\pi_\psi) - H(\pi_{\psi} \psi + \pi_\psi F + 4 \pi_F F)    \nonumber \\
&   &  \mbox{} + {4a^3\over{\N}}(H^2 - \dot{H} + H \nubble)(3\psi^2 - 2\psi F +
F^2)  + 2\N a\Delta\psi \psi \nonumber \\
&   &  \mbox{} + \sum_A \biggl( (\dot{\varphi}^A \dpfi - \N a^3 V^A_{,\varphi}
\delta \varphi _A)(3\psi -F) + {\N \over{2a^3}}(\dpfi)^2 \nonumber \\
&   &  \qquad - {\N a\over 2} \Delta\dphi^A\dphi_A + {\N a^3\over 2}V^A_{,
\varphi\varphi}(\dphi_A)^2 \biggr) ,
\end{eqnarray}
and we introduced the new independent variables\footnote{Here and throughout
the paper, $\Delta$ denotes the flat three-dimensional Laplacian, e.i., $\Delta
E \equiv E_{,ii}$.}

\begin{equation}
\begin{array}{rclrcl}
F &  = &  \Delta E, & & &  \nonumber   \\
\pi_F &  = &  2(\1s + \Delta\2s), &  \pi_\psi &  = & -2(3 \1s  + \Delta
\2s), \\
 \,^{\scriptscriptstyle (s)}\lambda^0 &  = & \dN,  &
\,^{\scriptscriptstyle (s)}\lambda^1 & = & \Delta B. \nonumber
\end{array}
\end{equation}

For the {\bf vector perturbations} one gets:
\begin{equation}
\,^{\scriptscriptstyle (v)}S = \int dt d^3x \biggl(\pi_V^i \dot{V}_i  -
\,^{\scriptscriptstyle (v)}\lambda^i \,^{\scriptscriptstyle (v)} C_i -
\,^{\scriptscriptstyle (v)} \H \biggr) \label{vectoraction}
\end{equation}
where
\begin{eqnarray}
\,^{\scriptscriptstyle (v)} C_i & = & \pi_V^i + {4 H a^3 \over \N} \Delta V^i
\nonumber \\
\,^{\scriptscriptstyle (v)} \H  & = & - {2 a^3\over \N} (H^2 - {\dot H} + H
\nubble)V^i \Delta V_i - 2H \pi_V^i V_i ~,
\end{eqnarray}
and we introduced instead of $\vpi^i$ and $D^i$ the independent variables
\begin{equation}
\pi_V^i = -2 \Delta \vpi^i, \qquad \,^{\scriptscriptstyle (v)}\lambda^i = D^i +
{\N \over a^3}\vpi^i
\end{equation}
Note that in this case we also have only two independent constraints, since
\begin{equation}
\,^{\scriptscriptstyle (v)}\lambda^i_{~,i} = 0~, \qquad \,^{\scriptscriptstyle
(v)} C^i_{~,i} = 0.
\end{equation}
For the {\bf tensor perturbations} there are no constraints and the action
takes the form
\begin{equation}
\,^{\scriptscriptstyle (t)}S = \int dt d^3x \biggl(\pi_h^{ij} {\dot h}_{ij}  -
\,^{\scriptscriptstyle (t)} \H \biggr), \label{tensorraction}
\end{equation}
where $\pi_h^{ij} \equiv \tpi^{ij}$ and
\begin{eqnarray}
\,^{\scriptscriptstyle (t)} \H & = &  {\N\over{a^3}}\pi_h^{ij}\pi_h^{ij} - 4H
\pi_h^{ij} h_{ij} \nonumber \\
 & & \mbox{}+ {a^3\over{\N}}(H^2 - \dot{H} + H\nubble)h_{ij}h_{ij}  - {\N
a\over{4}}\Delta h_{ij} h_{ij}.
\end{eqnarray}
Let us mention an interesting point. Starting with a superhamiltonian and
fixing the background, we end up with a pure Hamiltonian for perturbations
which is responsible for their dynamics  with respect to background time. There
are still four independent constraints (two for scalar perturbations plus two
for vector ones), as it should be. These constraints are linear in momenta and
generate themselves the gauge transformations which correspond to
diffeomorphisms \cite{Mukh_Wipf}.

\section{Gauge-invariant variables}

The constraints $\,^{\scriptscriptstyle (s)} C_0, \,^{\scriptscriptstyle (s)}
C_1, \,^{\scriptscriptstyle (v)} C_i $ and the Hamiltonian
$$\H(t)= \int d^3x(\,^{\scriptscriptstyle (s)}\H + \,^{\scriptscriptstyle
(v)}\H+ \,^{\scriptscriptstyle (t)}\H )
$$
form a closed algebra with respect to Poisson brackets $\{..,..\}$ defined in
the standart manner for our canonically conjugated variables $\dphi^a, \dpfi;
\psi, \pi_\psi; F,$ $\pi_F; V_i, \pi_V^i; h_{ij},\pi_h^{ij}$;
that is
\begin{equation}
\biggl\{C_\alpha, C_\beta\biggr \} = t_{\alpha \beta}^\gamma C_\gamma \qquad
\mbox{and} \qquad   \biggl\{\H(t), C_\alpha\biggr \} = t_\alpha^\beta C_\beta +
{\partial C_\alpha \over{ \partial t}}. \label{strcoeff}
\end{equation}
For instance, for the {\it scalar} constraints $\,^{\scriptscriptstyle
(s)}C_0,\,^{\scriptscriptstyle (s)}C_1$ we have \footnote{The appearance of
time derivatives on the right-hand-side is due to the explicit time dependence
of the constraints.}
\begin{eqnarray}
\biggl\{ C_0(x,t),C_0(y,t)\biggr \} & = & 0=\biggl\{ C_1(x,t),C_1(y,t)\biggr
\},  \nonumber \\
\biggl\{\H(t), C_0(x,t), \biggr \} & = & {\N\over{a^2}}\Delta_x C_1(x,t) +
{\partial C_0(x,t)\over{\partial t}},  \label{scalarcoeff} \\
\biggl\{\H(t), C_1(x,t) \biggr \} & = &  {\partial C_1(x,t)\over{\partial t}}.
\nonumber
\end{eqnarray}
The action~(\ref{scalaraction}) is invariant  with respect to the
transformations generated by the constraints~(\ref{scalarconstr}), if we
simultaneously transform the Lagrangean multipliers $\,^{\scriptscriptstyle
(s)}\lambda^{\alpha}$:
\begin{equation}
\begin{array}{lll}
\delta_{\xi} q & = & \{ q, \xi^{\alpha} C_{\alpha} \}  \\
\delta_{\xi} \lambda^{\alpha} & = & \dot{\xi}^{\alpha} - \xi^{\beta}
\lambda^{\gamma}t^\alpha_{\gamma \beta} - \xi^\beta t^\alpha_\beta
\label{transfo}
\end{array}
\end{equation}
where $\xi \equiv \xi^\alpha$ are the parameters of transformation and $q$ can
be any of the canonical variables $\dphi^a, \dpfi, \psi$, etc.

Since the constraints are linear in momenta,  the
transformations~(\ref{transfo}) correspond to diffeomorphisms \cite{Mukh_Wipf}.
{}From~(\ref{transfo}) and~(\ref{scalaraction}) with taking into
account~(\ref{strcoeff}),~(\ref{scalarcoeff}), it is easy to get the following
transformation laws:
\begin{equation}
\begin{array}{rclrcl}
\delta_{\xi} (\dphi_A) &  = & {\dot \varphi}_A \xi^0/ \N , & \delta_{\xi} \psi
& = & -H \xi^0/ \N ,  \nonumber \\
\delta_{\xi} \lambda^1 & = & {\dot \xi}^1 - \N \Delta \xi^0/a^2 , &
\delta_{\xi} \lambda^0 & = & {\dot \xi}^0 , \label{scalardiffeo} \\
\delta_{\xi} F & = & \xi^1 , & \mbox{etc} & &
\end{array}
\end{equation}
Using~(\ref{scalardiffeo}) one can easily construct gauge invariant variables
$Q$ for which $\delta_\xi Q = 0 $. For instance, the most important of them
are:
\begin{eqnarray}
v_A & = & a(\delta\varphi_A + {\dot{\varphi}_A\over{H}}\psi) \label{givariable}
\\
\Delta \Phi & = & {1\over{a}}\Delta \lambda^0 - {1\over{a}}{\partial \over
\partial t}[{a^2\over{\N}}(\lambda^1 - \dot{F})]. \label{Newton}
\end{eqnarray}
As we will see, the first variable naturally appears when we quantize the
theory, whereas the second one has clear physical interpretion, namely, $\Phi$
corresponds to the relativistic Newtonian potential \cite{MFB92}.

The gauge invariant variables for the vector perturbations can be constructed
in the same manner. The tensors $h_{ij}$ and $\pi_h^{ij}$ are gauge invariant
by themselves since there are no constraints which contain these variables.
For more details, refer to \cite{And_Mukh}.

\section{Path integral}

To quantize the perturbations we write the path integral ($\hbar = 1$)
\begin{equation}
K = \int_i^f {\cal D}\mu e^{i(^{\scriptscriptstyle (s)}S + ^{\scriptscriptstyle
(v)} S + ^{\scriptscriptstyle (t)} S)}
\end{equation}
which is just the propagator from some initial $(i) $ to final $(f)$ state. The
diffeomorphism invariant measure ${\cal D}\mu$ can be taken as
\begin{equation}
{\cal D}\mu = {\cal D}(\delta\gamma_{ij})  {\cal D}(\delta \pi^{ij}) {\cal
D}(\delta \N^{\alpha}) = {\cal J}\,{\cal D}^{\scriptscriptstyle (s)}\mu {\cal
D}^{\scriptscriptstyle (v)}\mu {\cal D}^{\scriptscriptstyle (t)}\mu
\end{equation}
where ${\cal J}$ is an irrelevant constant Jacobian which can be absorbed in
normalization and
\begin{eqnarray}
{\cal D}^{\scriptscriptstyle (s)}\mu & = & {\cal D}(\dphi_A) {\cal D}(\dpfi)
{\cal D}\psi {\cal D}\pi_\psi {\cal D}F {\cal D}\pi_F {\cal
D}^{\scriptscriptstyle (s)}\lambda^0 {\cal D}^{\scriptscriptstyle
(s)}\lambda^1, \nonumber \\
{\cal D}^{\scriptscriptstyle (v)}\mu & = & {\cal D}V_i {\cal D}\pi_V^i {\cal
D}^{\scriptscriptstyle (v)}\lambda^i , \label{measure} \\
{\cal D}^{\scriptscriptstyle (t)}\mu & = & {\cal D}h_{ij} {\cal D}\pi_h^{ij}
\nonumber
\end{eqnarray}
are the measure, correspondingly, for scalar, vector and tensor perturbations.
For ${\cal D}(\dphi^A), {\cal D}(\dpfi), {\cal D}\psi, {\cal D}\pi_\psi$ etc,
we  take the usual Liouville measure. Then the propagator $K$ can be
represented as a product of propagators for different types of perturbations.
Let us calculate them separately.

For the {\bf scalar perturbations}:
\begin{equation}
^{\scriptscriptstyle (s)}K = \int {\cal D}^{\scriptscriptstyle (s)}\mu
e^{i^{\scriptscriptstyle (s)}S}
\end{equation}
where $^{\scriptscriptstyle (s)}S$ is given by~(\ref{scalaraction}) and ${\cal
D}^{\scriptscriptstyle (s)}\mu $ is defined in~(\ref{measure}).

Let us integrate at first over the Lagrangean multipliers $^{\scriptscriptstyle
(v)}\lambda^0, ^{\scriptscriptstyle (v)}\lambda^1$. Then we get the delta
functions of constraints $^{\scriptscriptstyle (s)} C_\alpha$, and since they
are linear in momenta they permit us to integrate easaly over $\pi_\psi,
\pi_F$. Then, changing the variables (in particular, introducing the gauge
invariant variables $v_A$, see~(\ref{givariable}), instead of $\dphi_A$), with
help of the eqs. of motion for background~(\ref{eq:velocities}),~(\ref{KGeqn})
and~(\ref{eq:background}) and after integration over the momenta we obtain the
following result\footnote{We also skip from the action a lot of total
derivatives.}
\begin{equation}
^{\scriptscriptstyle (s)}K = {\cal M} \int \prod_A  {\cal D} v_A
e^{i^{\scriptscriptstyle (s)}S(v_A)}~,  \qquad  {\cal D} v_A = \prod_i {dv_i^A
\over \sqrt{2 \pi i \Delta \eta}}
\end{equation}
where
\begin{equation}
{\cal M} \propto \int {\cal D}\psi {\cal D}F \label{gaugevol}
\end{equation}
corresponds to the explicitly factorized infinite volume of the gauge group
and can be absorbed in normalization. The action $ S(v_A)$ depends only on the
gauge invariant variable $v_A$, (see~(\ref{givariable})):
\begin{equation}
\,^{\scriptscriptstyle (s)} S(v_A) = {1 \over 2} \int d^3x d\eta \biggl (\sum_A
(v'^2_A - v^A_{,i} v^A_{,i}) + \sum_{A,B} \Omega_{AB} v^A v^B \biggr ).
\label{vaction}
\end{equation}
Here the prime means derivative with respect to the new time parameter
\begin{equation}
\eta = \int {\N dt \over a},
\end{equation}
and the function of time $\Omega_{AB}$ is
\begin{eqnarray}
\Omega_{AB} & = & \biggl[ \biggl( {z_A'' \over z_A} + 2 ({\Ht' \over \Ht} -
\Ht){z_A' \over z_A} + {\Ht'' \over \Ht} - 2 \Ht^2 \biggr) \delta_{AB} + {1
\over 2 a^2} ( \Ht z_A z_B)'\biggr] ,\nonumber \\
& & z_A   =  {a \varphi_A' \over \Ht}, \qquad \Ht = {a' \over a}.
\end{eqnarray}
We see that the metric perturbations $\psi, F$ do not have their own dynamical
degrees of freedom. They are entirely due to the perturbations in the matter
distribution. In this sense the quantization of scalar cosmological
perturbations is just the same as quantization of matter itself with taking
into account the gravitational field created by this matter.

The {\bf vector perturbations} do not possess any dynamics. This can easily be
seen if we integrate in $^{\scriptscriptstyle (v)} S$, at first over the
Lagrangean multiplier $^{\scriptscriptstyle (v)} \lambda^i$ and then over the
momenta $\pi_V^i$. As a result we find that the action $^{\scriptscriptstyle
(v)} S(V_i)$ vanishes (up to the total derivatives). The integral over $V_i$
corresponds to the integration over the gauge volume, like in~(\ref{gaugevol}).

If the Universe would be filled by some matter which has vector degrees of
freedom (like, for instance, by a perfect fluid) then the quantization of
vector perturbations would become nontrivial.

For {\bf tensor perturbations} the calculations are very simple. Integrating
over the momenta $\pi_h^{ij}$ one arrives at:
\begin{equation}
^{\scriptscriptstyle (t)} K = \int {\cal D}h_{ij} {\cal D}\pi_h^{ij} e^{i
\,^{\scriptscriptstyle (t)} S}  = \int {\cal D}e_{ij} e^{i
\,^{\scriptscriptstyle (t)} S(e_{ij})}~,
\end{equation}
where
\begin{equation}
\,^{\scriptscriptstyle (t)} S = {1\over 2} \int d\eta d^3x [e'_{ij} e'_{ij} -
e_{ij,m} e_{ij,m} + {a''\over a} e_{ij} e_{ij}]~,
\end{equation}
and $e_{ij} = a h_{ij}/ \sqrt{2}$ is the the gauge invariant variable.

Thus the quantization of the cosmological perturbations in a Friedmann
Universe has been reduced to the quantization of a set of gauge invariant
fields with time-dependent masses in a flat space-time.

\section{Propagator}

As an example, we calculate the propagator and the evolution of an initial
vacuum state for scalar perturbations in the Schr\"{o}dinger representation. To
simplify the consideration we assume that there is only {\it one} scalar field
in the Universe, that is $v_A \equiv v$ and $\Omega_{AB} \equiv \Omega =
z''/z$.

Expanding $v(\vec{x},\eta)$ in Fourier modes as
\begin{equation}
v(\vec{x},\eta) = \sqrt{2} \biggl( \int \limits_{\scriptscriptstyle k_3>0}
{d^3k \over (2\pi)^{3/2} }v_k(\eta) \sin(\vec{k} \vec{x}) +  \int
\limits_{\scriptscriptstyle k_3\leq 0} {d^3k \over (2\pi)^{3/2} }v_k(\eta)
\cos(\vec{k} \vec{x}) \biggr)
\end{equation}
where the Fourier coefficients $v_k$ are real, one gets
\begin{equation}
 S(v) = {1 \over 2} \int d^3k d\eta  \biggl(v_k'^2 - k^2 v_k^2 + \Omega(\eta)
v_k^2 \biggr). \label{QM_action}
\end{equation}
Thus, we can write the propagator from some initial $(i)$ to final $(f)$ field
configuration as
\begin{equation}
K(f|i) \propto \prod_k K_k(v_f, \eta_f | v_i, \eta_i),
\end{equation}
where\footnote{To simplify the notation we also skip the index  $k$ in some of
the formulae whenever it is clear which of functions
depend on $k$.}
\begin{equation}
K_k (v_f,\eta_f | v_i, \eta_i) = \int \limits_{\scriptscriptstyle
v_k(\eta_i)=v_i}^{\scriptscriptstyle v_k(\eta_f)=v_f} {\cal D}v_k \exp \biggl\{
{i\over 2} \int \limits_{\scriptscriptstyle \eta_i}^{\scriptscriptstyle \eta_f}
d\eta (v_k'^2 - k^2 v_k^2 + \Omega(\eta) v_k^2) \biggr\} \label{propagator}
\end{equation}
is the propagator for the harmonic oscillator with the time-dependent frequency
\begin{equation}
\omega_k^2(\eta) = k^2 - \Omega(\eta). \label{omega}
\end{equation}

The result for the path integral~(\ref{propagator}) is well known
\cite{Kleinert, Feyn_Hibbs} :
\begin{equation}
K_k(v_f,\eta_f|v_i, \eta_i) = \sqrt{{-1\over 2\pi i} {\partial^2 S_{cl} \over
\partial v_f \partial v_i }}\, e^{i S_{cl}[v_f,v_i]},
\end{equation}
where $S_{cl}$ is the action for the classical trajectory satisfying the
equation
\begin{equation}
v_k'' + \omega^2_k v_k = 0 \label{v_EoM}
\end{equation}
and the boundary conditions $v_k(\eta_i) = v_i, v_k(\eta_f) = v_f$.

The classical action $S_{cl}$ can be expressed as a combination of two
fundamental solutions $C(\eta),D(\eta)$ of the eq.~(\ref{v_EoM}):
\begin{equation}
v_k(\eta) = v_i C(\eta) + v_f D(\eta)
\end{equation}
Here the functions $C$ and $D$ depend on $k$, but not on $v_i,v_f$ and they
satisfy the boundary conditions
\begin{equation}
\begin{array}{cc}
C(\eta_i) = 1~, & C(\eta_f) = 0~, \nonumber \\
D(\eta_i) = 0~, & D(\eta_f) = 1~. \label{bound_cond}
\end{array}
\end{equation}
In terms of $C$ and $D$, the explicit expression for the
propagator~(\ref{propagator}) is given by:
\begin{equation}
K_k(v_f,\eta_f|v_i, \eta_i) = \biggl({1\over 2 \pi i
f(\eta_f,\eta_i)}\biggr)^{1/2} \exp\biggl\{{i\over 2} (D'_fv_f^2 - C'_i v_i^2 -
2 {v_i v_f \over f(\eta_f,\eta_i)})\biggr\}
\end{equation}
where
\begin{equation}
f(\eta_f,\eta_i) = {2\over D'_i- C'_f} \qquad \mbox{and} \qquad C'_i \equiv
{\partial \over \partial \eta}C(\eta_i)~, \mbox{etc}
\end{equation}
do not depend of $v_i, v_f$.

Given the propagator, we can calculate the evolution of any initial state
functional
$\Psi(v_i,\eta_i)$. For instance, if this initial state can be represented in
the form
\begin{equation}
\Psi \biggl(v_i(\vec{x},\eta_i),\eta\biggr) \propto \prod_k
\Psi_k(v_i,\eta_i)~,
\end{equation}
then as a result of evolution we find that
\begin{equation}
\Psi \biggl(v_f(\vec{x},\eta_f),\eta \biggr) \propto \prod_k \Psi_k(v_f,\eta_f)
\end{equation}
where
\begin{equation}
\Psi_k(v_f,\eta_f) = \int \limits_{- \infty}^{+  \infty} dv_i K_k(v_f,\eta_f |
v_i, \eta_i) \Psi_k(v_i,\eta_i) .\label{evolution}
\end{equation}
Thus, there will be no mixing of the Fourier components with different $k$ in
the linear approximation as it should be.

If we choose, for instance, a normalized gaussian state with dispersion
$\sigma_0$ (which can depend on $k$)  as initial wavefunction,
\begin{equation}
\Psi_k(v_i,\eta_i) = (\pi\sigma^2_0)^{-1/4} \exp \{-{v_i^2\over 2
\sigma_0^2}\} ,\label{vacuum}
\end{equation}
then the result for the integral~(\ref{evolution}) is
\begin{equation}
\Psi_k(v_f,\eta_f) = e^{{i\over 2}\theta} (\pi\Sigma^2)^{-1/4} \exp\biggl\{-
{1\over 2}v_f^2( {1\over\Sigma^2} - i\Lambda) \biggr\} ,  \label{statefunction}
\end{equation}
where $\Sigma, \Lambda$ and $\theta$ depend on $\eta_i,\eta_f,k$ and
$\sigma_0$:
\begin{eqnarray}
\Sigma^2 & \equiv  & {4\over \sigma_0^2}\biggl[{1+ \sigma_0^4 C_i'^2 \over
(D_i' -
C_f')^2} \biggr] \label{sigma}~, \qquad \Lambda  \equiv  D'_f + {\sigma_0^2
\over
\Sigma^2} C'_i \label{sigma_lambda} \\
\cot \theta & \equiv & \sigma_0^2 C'_i . \nonumber
\end{eqnarray}
Thus, the wave function at time $\eta_f$ describes still a gaussian state, but
now with a modified complex dispersion.

Since all information about the system is contained in the state
function~(\ref{statefunction}), it is standard to compute expection values
$\langle \hat{v}_k \hat{v}_k \rangle,$ etc.

If we want to take as initial state the vacuum one, we need to specify the
dispersion $\sigma_0$ entering~(\ref{vacuum}). As it is well known there is an
ambiguity in the definition of vacuum in the presence of external fields
\cite{Birrell_Davies}. In our case this ambiguity is due the time dependence of
the frequency $\omega_k$ in~(\ref{omega}). However for any definition of vacuum
we should have
\begin{equation}
\sigma_0^2 \longrightarrow {1\over k} \qquad \mbox{for} \qquad k^2 \gg
\Omega(k) \label{k_asym}
\end{equation}
Later we shall see that this condition is sufficient to get unambiguous
predictions for the spectrum of fluctuations in the most interesting range of
scales in inflationary model.

The final step is to calculate the two-point correlation function of the gauge
invariant relativistic potential $\Phi(\vec{x},\eta)$, (see~(\ref{Newton})).

\section{Correlation function and power spectrum}

For the purpose of comparing with observation we need the two-point
correlation function $\xi(r)$\footnote{In the case under consideration the
correlation function just depends on $r = |\vec{r}|$.} of the gauge invariant
metric fluctuation $\Phi(\vec{x},\eta)$:
\begin{eqnarray}
\xi(r) & = & \langle \Psi(v,\eta)|\hat{\Phi}(\vec{x})
\hat{\Phi}(\vec{x}+\vec{r})| \Psi(v,\eta) \rangle  \label{corr_function} \\
& = & \int\limits_0^\infty {dk\over k}{\sin (kr)\over kr} |\delta_k|^2
\label{power}
\end{eqnarray}
where $\Psi(v,\eta)$ is the state functional and the power spectrum
$|\delta_k|$ is the  measure of the amplitude of metric fluctuations on a
comoving scale $\sim 1/k$.
The relation between $\Phi(\vec{x},\eta)$ and $v(\vec{x},\eta)$, and,
correspondingly, between the quantum operators $\hat{\Phi}$ and $\hat{v}$,
follows from Hamilton eqs., (see \cite{MFB92}):
\begin{equation}
\Delta \Phi = {\varphi'\over 4 a}(\pi_v - {z'\over z}v) \label{v_phi}
\end{equation}
where $\pi_v$ is the momentum conjugated to $v$.

In the Schr\"{o}dinger representation, working with separate Fourier
modes, one has for the appropriate momentum  operator $\hat{\pi}_k =  -i
\partial/ \partial v_k$.

Now using the definitions~(\ref{corr_function}), taking into
account~(\ref{power}) and~(\ref{v_phi}) and calculating the corresponding
correlation functions
\begin{equation}
\langle \hat{v}_k \hat{v}_k \rangle~, \qquad \langle
\hat{\pi}_k \hat{\pi}_k \rangle~, \qquad \langle \hat{v}_k \hat{\pi}_k +
\hat{\pi}_k
\hat{v}_k \rangle~,
\end{equation}
for the state~(\ref{statefunction}), we get the following result for
the power spectrum $|\delta_k|$:
\begin{eqnarray}
|\delta_k|^2 &  = & {4\pi \over (2\pi)^3} k^3 \langle \hat{\phi}_k \hat{\phi}_k
\rangle \nonumber \\
 & = & ({\varphi' \over 8 \pi a})^2 {1\over k} \biggl[{1\over
\Sigma^2} +  \Sigma^2 (\Lambda - {z'\over z})^2   \biggr] \label{spectrum}
\end{eqnarray}

Thus, given an initial gaussian state at $\eta = \eta_i$, specified by
the dispersion $\sigma_0(k)$, we derived the power spectrum $|\delta_k|$ at
an arbitrary later moment of time $\eta = \eta_f$. This spectrum can be
expressed entirely in terms of the two functions $C(\eta)$ and $D(\eta)$
which satisfy the eq.~(\ref{v_EoM}) with the boundary
conditions~(\ref{bound_cond}).

The functions $\Sigma$ and $\Lambda$, in terms of $C$ and $D$, are given in
{}~(\ref{sigma_lambda}). For an initial ``vacuum'' state the asymptotic
dependence of $\sigma_0(k)$ on $k$ at $k^2 \gg \Omega$ is specified
in~(\ref{k_asym}).

As an example, let us consider chaotic inflation in the model with a  massive
scalar field: $V(\varphi) = {1\over 2}m^2 \varphi^2$.

In this case from
{}~(\ref{spectrum}) one obtains the following spectrum for the metric
fluctuation after the end of inflation in the most interesting range of
galactic scales:
\begin{equation}
|\delta_k| \simeq (\sqrt{3}/ 10 \pi)m \ln (\lambda_{ph}/\lambda_\gamma),
\end{equation}
where $\lambda_{ph} = a(\eta)/k$ is the physical wavelength of the
perturbation and $\lambda_\gamma$ is some characteristic wavelength of the
cosmic background radiation.
This result coincides with the one obtained by other methods (see
\cite{MFB92}). The inflation amplifies the initial vacuum fluctuations enough
to explain the large-scale structure of the Universe only if $m\sim 10^{13}
GeV$. The other interesting applications are considered in a forecoming paper
\cite{And_Mukh}.

\section{Discussion}

Starting with the Hamiltonian (ADM) formulation of General Relativity, we
deduced the Hamiltonian theory of cosmological perturbations in the Friedmann
Universe.

To derive the action for linearized metric and matter perturbations in a first
order Hamiltonian formalism, we expanded the ADM action for gravity and the
action for the matter (scalar fields) up to second order in perturbations. The
concrete calculations have been done for the case when the background
Friedmann Universe has zero spatial curvature. This is a good approximation
if, afterwards, we want to quantize only the perturbations assuming that the
background is classical.

The generalization of the developed formalism to the closed Universe is
straightforward and will be presented in the future publication
\cite{And_Mukh}. This case is especially interesting if we also want to
quantize the Universe as a whole.
However if the zero mode perturbation which correspond to the Friedmann
background is in a quasiclassical region, then only the quantization of
inhomogeneities becomes interesting.

We have further studied the diffeomorphism transformations in Hamiltonian
formalism
and constructed explicitly gauge (diffeomorphism) invariant variables for
perturbations.

Then, the first order Hamiltonian theory was quantized via well defined
Hamiltonian path integral. The volume of the gauge group (diffeomorphism) was
factorized explicitly  and the problem was reduced to the quantization of a
set of gauge invariant fields and, finally, to the quantization of harmonic
oscillators with time dependent frequences.

We calculated the propagator which describes the evolution of the quantum
state and then found a closed expression for the power spectrum of
fluctuations for initial gaussian, for instance, vacuum state.

All the calculations have been done in the Schr\"{o}dinger picture which is
very convenient to study, for example, the decoherence problem for the
cosmological perturbations \cite{Kiefer,Halliwell}. The results were finally
applied to the concrete model of chaotic inflation.

{\bf Acknowledgments:} We thank C. Schmid and A.Wipf for valuable discussions.
This work has been supported in parts by the Swiss National Science Foundation.

\end{document}